\documentclass[apj]{emulateapj}
\usepackage{apjfonts}
\shortauthors{Tran et al.}
\shorttitle{Reversal of the Star Formation-Density Relation in a
Cluster at $z=1.62$}

\begin{document}

\newcommand{\mipsmu}{$24\mu$m}
\newcommand{\msunyr}{~M$_{\odot}$~yr$^{-1}$}
\newcommand{\sfrsed}{5\msunyr}
\newcommand{\irc}{ClG J0218.3-0510}

\title{Reversal of Fortune: Confirmation of an \\Increasing Star
Formation-Density Relation in a Cluster at $z=1.62$\altaffilmark{1}}

\author{Kim-Vy H. Tran\altaffilmark{2,3}, 
Casey Papovich\altaffilmark{2}, 
Am\'elie Saintonge\altaffilmark{4}, 
Mark Brodwin\altaffilmark{5},
James S. Dunlop\altaffilmark{6},
Duncan Farrah\altaffilmark{7}, 
Keely D. Finkelstein\altaffilmark{2}, 
Steven L. Finkelstein\altaffilmark{2},
Jennifer Lotz\altaffilmark{8},
Ross J. McLure\altaffilmark{6},
Ivelina Momcheva\altaffilmark{9}, 
%Gregory Rudnick\altaffilmark{10}, 
\& Christopher N. A. Willmer\altaffilmark{10}} 

\altaffiltext{1}{This work is based in part on observations made with
the Spitzer Space Telescope, which is operated by the Jet Propulsion
laboratory, California Institute of Technology, under NASA contract
1407.  This paper also includes data gathered with the 6.5 meter
Magellan Telescopes located at Las Campanas Observatory, Chile.  This
work is based in part on data collected at Subaru Telescope, which is
operated by the National Astronomical Observatory of Japan.}
\altaffiltext{2}{George P. and Cynthia W. Mitchell Institute for
Fundamental Physics and Astronomy, Department of Physics,  
Texas A\&M University, College Station, TX 77843; vy@physics.tamu.edu}
\altaffiltext{3}{Institute for Theoretical Physics, University of
  Z\"urich, CH-8057 Z\"urich, Switzerland}
\altaffiltext{4}{Max-Planck-Institut f\"ur Extraterrestrische Physik,
Giessenbachstra\ss e, D-85748 Garching, Germany} 
\altaffiltext{5}{W. M. Keck Postdoctoral Fellow, Harvard--Smithsonian
Center for Astrophysics, 60 Garden St., Cambridge, MA 02138} 
\altaffiltext{6}{SUPA (Scottish Universities Physics Alliance),
Institute for Astronomy, University of Edinburgh,
Royal Observatory, Edinburgh, EH9 3HJ, UK}
%\altaffiltext{6}{Institute for Astronomy, Royal Observatory,
%University of Edinburgh, UK}
\altaffiltext{7}{Astronomy Centre, University of Sussex, Falmer,
Brighton, UK}
\altaffiltext{8}{Leo Goldberg Fellow, National Optical Astronomy Observatories,
 950 N.~Cherry Ave., Tucson, AZ 85719}
\altaffiltext{9}{Observatories, Carnegie Institution of Washington,
813 Santa Barbara Street, Pasadena, CA, 91101}
%\altaffiltext{10}{Department of Physics and Astronomy, University of Kansas, 1251 Wescoe Hall Dr., Lawrence, KS, 66045-7582}
\altaffiltext{10}{Steward Observatory, University of Arizona, 933
N.~Cherry Avenue, Tucson, AZ 85721}

\begin{abstract}

We measure the rest-frame colors (dust-corrected), infrared
luminosities, star formation rates, and stellar masses of 92 galaxies
in a Spitzer-selected cluster at $z=1.62$.  By fitting spectral energy
distributions (SEDs) to 10-band photometry
($0.4\mu$m$<\lambda_{obs}<8\mu$m) and measuring \mipsmu\ fluxes for
the 12 spectroscopically confirmed and 80 photometrically selected
members, we discover an exceptionally high level of star formation in
the cluster core of $\sim1700$\msunyr~Mpc$^{-2}$.  The cluster
galaxies define a strong blue sequence in $(U-V)$ color and span a
range in color.  We identify 17 members with
L$_{IR}>10^{11}$~L$_\odot$, and these IR luminous members follow the
same trend of increasing star formation with stellar mass that is
observed in the field at $z\sim2$.  Using rates derived from both the
\mipsmu\ imaging and SED fitting, we find that the relative fraction
of star-forming members triples from the lowest to highest galaxy
density regions, e.g. the IR luminous fraction increases from
$\sim8$\% at $\Sigma\sim10$~gal~Mpc$^{-2}$ to $\sim25$\% at
$\Sigma\gtrsim100$~gal~Mpc$^{-2}$.  The observed increase is a
reversal of the well-documented trend at $z<1$ and signals that we
have reached the epoch when massive cluster galaxies are still forming
a substantial fraction of their stars.

\end{abstract}


%%%%%%%%%%%%%%%%%%%%%%%%%%%%%%%%%%%%%%%%%%%%%%%%%%%%%%%%%%%%%%%%%%%%
\section{Introduction}

A well-established observational hallmark of how galaxies evolve as a
function of environment is the star formation-galaxy density relation.
A plethora of studies utilizing multi-wavelength tracers of activity
have shown that star formation universally decreases with increasing
galaxy density at $z<1$
\citep[e.g.][]{hashimoto:98,ellingson:01,gomez:03,patel:09}.  In
particular, the cores of massive galaxy clusters are galaxy graveyards
full of massive spheroidal systems that are dominated by old stellar
populations.  However, as we approach the epoch when these quiescent
behemoths should be forming the bulk of their stars
\citep[$z\gtrsim2$;][]{vandokkum:98b,jorgensen:06,rettura:10}, the
star formation-density relation should weaken and possibly reverse.
Identifying when star formation is quenched as a function of galaxy
mass and environment provides strong constraints on galaxy models
\citep[e.g.][]{kauffmann:93,hopkins:08}, i.e. is individual galaxy
mass or the larger scale environment the primary driver of evolution?

Observations of field galaxies at $z\sim1$ indicate that the star
formation-density relation turns over at this epoch such that there is
an enhancement of activity in the highest density regions of the field
\citep{elbaz:07,cooper:08}.  Studies also find an excess of dust-obscured star
formation in group environments at $z<1$
\citep{koyama:08,tran:09,gallazzi:09}, and very recent results suggest
that star formation may be enhanced in the significantly richer core
of a galaxy cluster at $z\sim1.46$ \citep{hilton:10}.
As cluster surveys push to
higher redshifts ($z>1$) and thus closer to the epoch when massive
galaxies are forming their stars, variations in age will become
evident in, e.g. a larger scatter in color, and robust star formation
rates should reveal increasing levels of activity even in the cluster
cores.

We report here the first confirmation of increasing star formation
activity with increasing galaxy density observed in cluster \irc1, a
Spitzer-selected galaxy cluster at $z=1.62$ \citep[hereafter
Pap10]{papovich:10}.  We use cosmological parameters $\Omega_m=0.3$,
$\Lambda=0.7$, and $H = 70$ km s$^{-1}$ Mpc$^{-1}$ throughout the
paper; at $z=1.62$, this corresponds to an angular scale $0.5$~Mpc
arcmin$^{-1}$.  All magnitudes are in the AB system.

%%%%%%%%%%%%%%%%%%%%%%%%%%%%%%%%%%%%%%%%%%%%%%%%%%%%%%%%%%%%%%%%%%%%
\section{Multi-wavelength Data}

\irc\ has a wealth of multi-wavelength imaging data that includes
$BRi^\prime z^\prime$ imaging from the Subaru-XMM Deep Survey
\citep{furusawa:08} and $JK$ imaging from the UKIRT IR Deep Sky Survey
\citep[UKIDSS,][]{lawrence:07}; for these data, we utilized the
$K$--selected catalog from \citet{williams:09}.  The cluster field
also has deep \textit{Spitzer} imaging available in the four IRAC
bands (3.6--8.0$\mu$m) and MIPS \mipsmu\ as part of the
\textit{Spitzer} public legacy survey of the UKIDSS Ultra Deep Survey
(SpUDS, PI:
J.~Dunlop)\footnote{http://ssc.spitzer.caltech.edu/spitzermission/observingprograms/legacy/spuds/}.
We matched the $K$--band selected catalog to the IRAC data following
\citet{papovich:06}.

\irc1 was originally detected as a strong overdensity of galaxies at
$1.5<z_{phot}<1.7$ with a mean surface density $>20\sigma$ higher than
that of galaxies in the same redshift range in the UKIDSS Ultra-Deep
Survey \citep{lawrence:07}.  Follow-up spectroscopy confirmed 10
members that define a narrow redshift peak at $1.622\leq z\leq 1.634$
with additional members at $z=1.609$ and $z=1.649$
\citep[Pap10;][]{tanaka:10}.  In our analysis, we include 80 more
members that are selected from photometric redshifts determined with
EAZY \citep{brammer:08}; we use the limit for the integrated redshift
probability defined by the spectroscopically confirmed members
($\mathcal{P}_z>0.3$; see Pap10 for a detailed description of the
photometric selection).

Analysis of XMM-Newton data in this field also reveals a weak detection
consistent with extended emission from the cluster.  The cluster mass
estimated from its X-ray luminosity is
$\sim1.1\times10^{14}$~M$_{\odot}$ and consistent with its estimated
dynamical mass of $\sim4\times10^{14}$~M$_{\odot}$.  To define the
cluster center, we select the massive (spectroscopically confirmed)
member located at the peak of the X-ray emission: its J2000
coordinates are (2:18:21.07, -5:10:32.84), and all of the cluster
galaxies lie within $R_{proj}\sim1$~Mpc of this member.

\subsection{Spectral Energy Distributions}

%However, we tested additional metallicities ranging from
%0.4--2.5~$Z_\odot$ and find that this assumption does not affect our
%conclusions.

We fit the 10--band galaxy photometry
($0.4\mu$m$<\lambda_{obs}<8\mu$m)\footnote{At $z=1.62$, this
corresponds to a rest-frame wavelength range of
$0.15\mu$m$<\lambda_{rest}<3.0\mu$m.} with the 2007 version of the
\citet{bruzual:03} stellar population synthesis models using a
Chabrier initial mass function \citep[for more details,
see][]{papovich:06}.  We find that models with Solar metallicity best
reproduce the rest-frame colors and scatter of the cluster red
sequence galaxies (Pap10), and we allow the models to range in age
from $10^6$ to $2\times 10^{10}$~yr; we also include dust attenuation
using the \citet{calzetti:00} law with color excess values ranging
from $E(B-V) = 0.0-0.7$.  We allow for a range of star formation
histories parameterized as a decaying exponential with an $e$--folding
time $\tau$, where at any age $t$ the star formation rate is $\Psi(t)
\sim \exp(-t/\tau)$ and $\tau$ ranges from 1 Myr to 100 Gyr
(corresponding approximately to instantaneous bursts to constant star
formation, respectively).  In the model fitting, we add a
$\sigma/f_\nu = 5$\% error in quadrature to the photometric
uncertainties on each band to account for mismatches in the multi-band
photometry, and for the fact that the models do not continuously
sample the model parameter space.  Our sample of 92 cluster members
has an average best-fit age of $\sim500$ Myr and an average best-fit
$\tau$ of $\sim500$ Myr; 35 members have SED-derived star formation
rates $>$\sfrsed.

\subsection{MIPS \mipsmu\ Fluxes}

The cluster field was imaged with Spitzer MIPS as part of the legacy
UKIDSS Ultra Deep Survey (SpUDS PI: J.~Dunlop).  We extracted sources
from the public SpUDS \mipsmu\ map using StarFinder
\citep{diolaiti:00}, an IDL-based PSF-fitting code designed for
crowded fields.  The \mipsmu\ image has 1.245\arcsec pixel$^{-1}$,
and we derived a model PSF and aperture corrections from the brightest
isolated sources from the SpUDS image.  The catalog includes all
sources detected with $S/N>5$; this corresponds to a flux of
$\sim40\mu$Jy.  Using simulated sources based on the PSF and injected
into the map, we determine that the catalog is 80\% complete at this
flux level.

The measured \mipsmu\ fluxes are converted into total infrared
luminosities (L$_{IR}$) using the \citet{chary:01} templates.  Recent
Herschel studies indicate that while this technique is very accurate
at $z<1.5$, extrapolations from monochromatic \mipsmu\ fluxes
overestimate the true $L_{IR}$ by factors of $2-7$ at $z>1.5$
\citep[e.g.][]{nordon:10}. Finally, star formation rates are
calculated from $L_{IR}$ using the prescription of
\citet{kennicutt:98}, adjusted to the Chabrier IMF.

% use (U-V) AB for easy comparison to Brammer+09
\begin{figure}
\epsscale{1.0}
\plotone{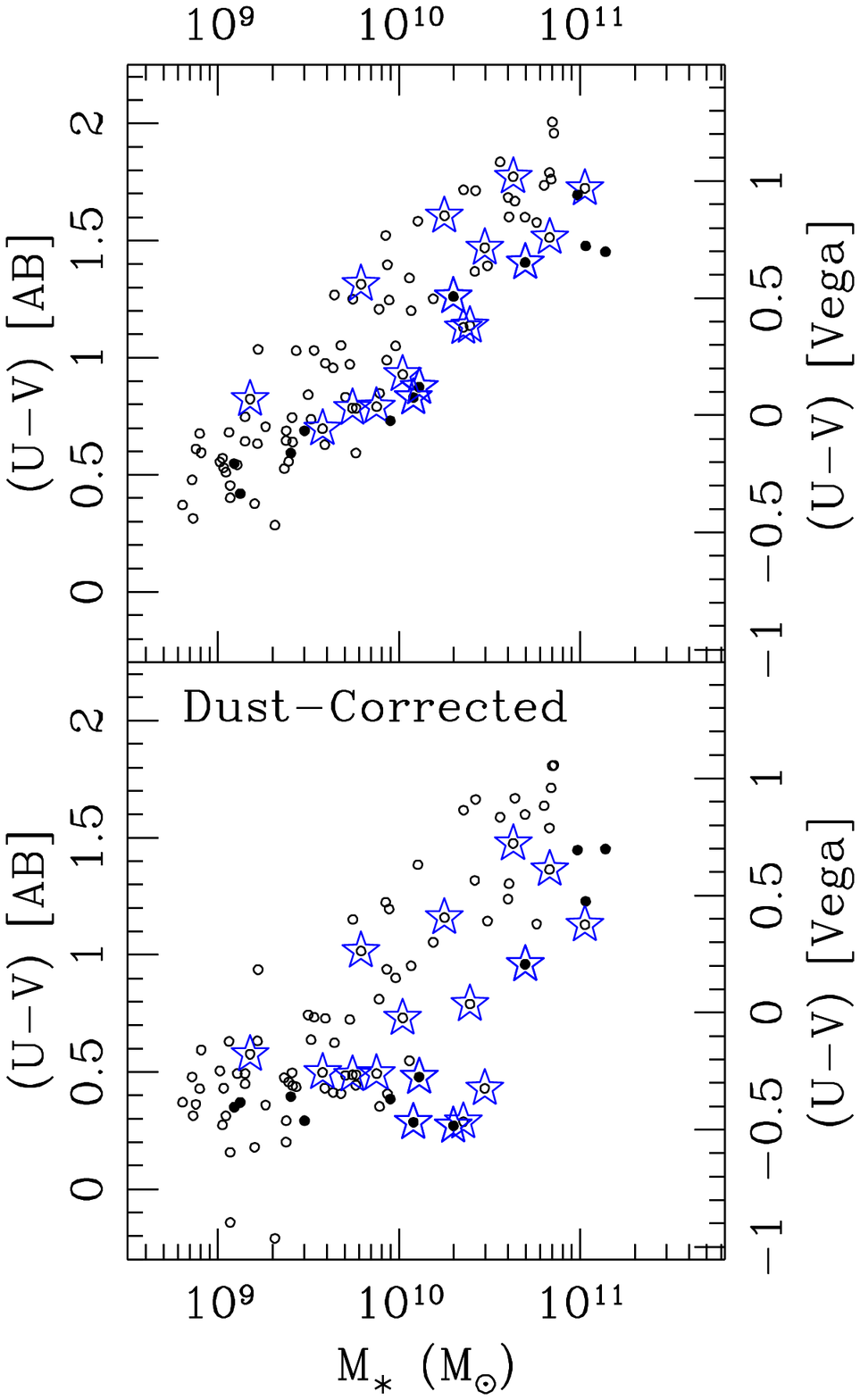}
\caption{Rest-frame $(U-V)_{AB}$ color vs. stellar mass determined
from fitting spectral energy distributions (SEDs; assuming Chabrier
IMF) to the 10-band photometry: shown are the measured ({\it top})
and dust-corrected ({\it bottom}) color-mass diagrams.  Open
circles denote  members selected using photometric redshifts from EAZY
\citep{brammer:08} and filled circles spectroscopically 
confirmed members from \citet{papovich:10} and \citet{tanaka:10}; the
latter tend to be on the blue edge because the spectroscopy favors
members with emission lines.
The 17 cluster members detected at \mipsmu\ are marked with open stars; 
note the number of IR luminous members that remain red even after being
corrected for dust extinction.  The \irc \ members have a color
distribution similar to that observed in the field at $z\sim2$
\citep{brammer:09} and, in contrast to galaxy clusters at $z<1$,
the members populate a {\it blue} sequence and span a range in
color. \label{fig:col_smass}}  
\end{figure}

%%%%%%%%%%%%%%%%%%%%%%%%%%%%%%%%%%%%%%%%%%%%%%%%%%%%%%%%%%%%%%%%%%%%
\section{Stellar Masses, Star Formation Rates, and Environment}

By fitting SEDs to the 10-band photometry
($0.4\mu$m$<\lambda_{obs}<8\mu$m), we are able to measure accurate
rest-frame colors (AB system) and stellar masses as well as correct
for dust extinction.  Figure~\ref{fig:col_smass} shows the measured
and dust-corrected rest-frame $(U-V)$ color versus stellar mass for
the 92 cluster galaxies within $R_{proj}\sim1$~Mpc of the BCG.  As
demonstrated by, e.g. \citet{wyder:07}, the galaxies begin to separate
into the well-known bimodal distribution only when the colors are
corrected for extinction.

The \irc \ members differ from their counterparts in clusters at
$z\lesssim1.2$ \citep[e.g.][]{holden:06,rettura:10} in that they have
a color-stellar mass distribution that is surprisingly similar to that
observed in the field at $z\sim2$ \citep{brammer:09}: the \irc \
members define a strong {\it blue} sequence and span a range in
(dust-corrected) color (Fig. 2, bottom panel).  While these members
populate a red sequence in the $(z-J)$ color-magnitude diagram
(Pap10), the correlation between rest-frame $(U-V)$ color and stellar
mass is visibly weaker\footnote{The astute reader will notice that the
red sequence is tighter in $(z-J)$ color because these filters
correspond approximately to rest-frame $(U-B)$ at $z=1.62$.}.
However, the most massive galaxies (M$_{\ast}\sim10^{11}$~M$_{\odot}$)
are still the reddest, i.e. they have the oldest stellar populations.

\begin{figure}
\epsscale{1.0}
\plotone{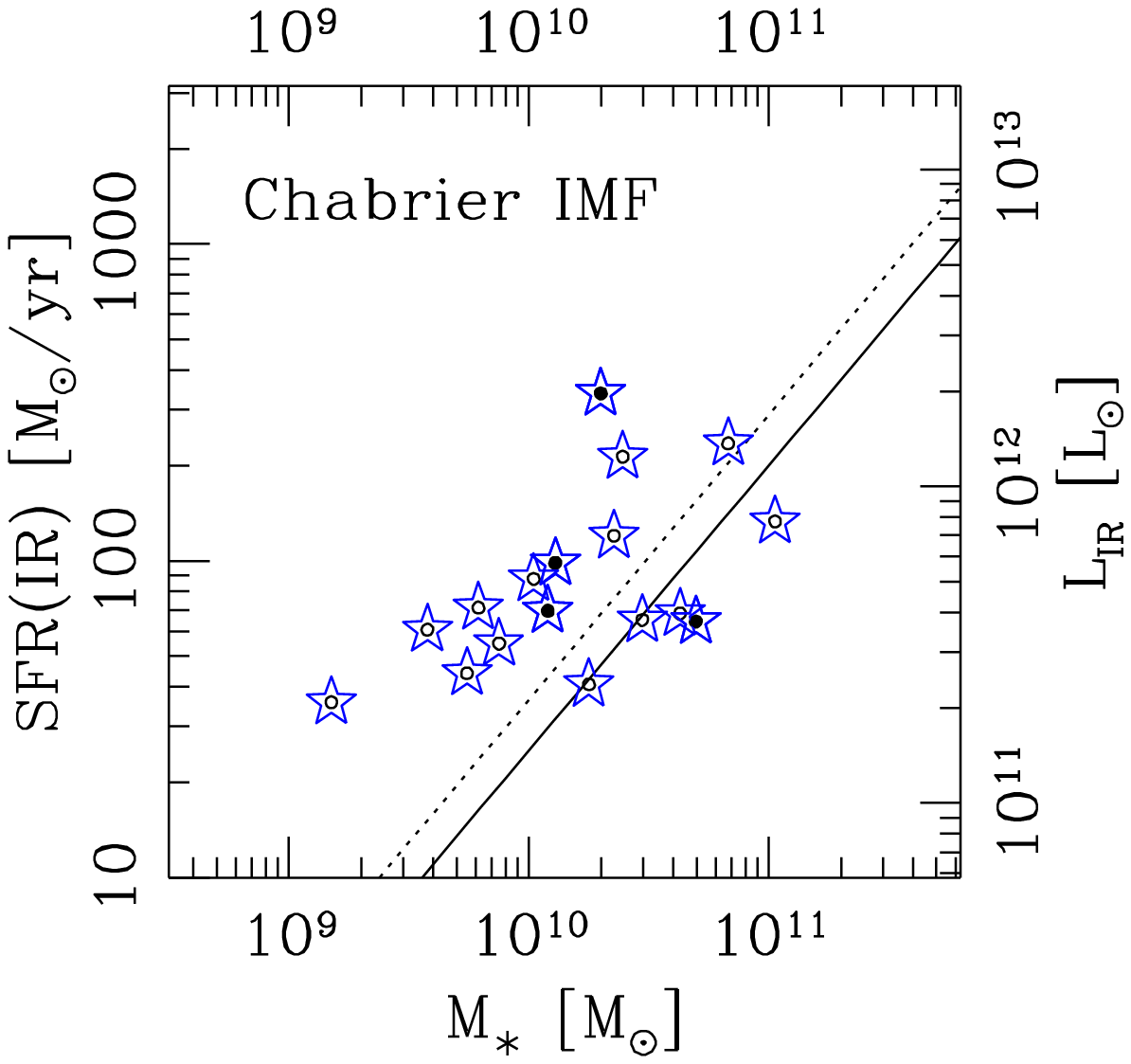}
\caption{Star formation rate vs. stellar mass for \mipsmu\ detected
cluster galaxies at $z=1.62$; symbols are as in
Fig.~\ref{fig:col_smass}.  {\it The most strongly 
star-forming systems are also some of the most massive cluster members.} 
These galaxies follow the same trend that is observed in field
galaxies at $z\sim2$ \citep[fitted relation and upper
semi-interquartile range of 0.16 dex shown as solid and dotted lines,
respectively;][]{daddi:07}.  All 17 of the \mipsmu\ detected members
have L$_{IR}>10^{11}$~L$_{\odot}$, in stark contrast to the relative
dearth of such massive, IR-bright galaxies in clusters at
$z<1$. \label{fig:sfr_smass}}  
\end{figure}

Seventeen of the members are detected at \mipsmu\
(Fig.~\ref{fig:col_smass}), and the most IR luminous members include
some of the most massive cluster galaxies (Fig.~\ref{fig:sfr_smass}).
Figure~\ref{fig:sfr_smass} shows that these IR luminous members follow
the same trend of increasing star formation with stellar mass that is
observed in the field at $z\sim2$ by \citet{daddi:07}.  Note that
because we can only detect members that are IR-bright
(L$_{IR}\gtrsim3\times10^{11}$~L$_{\odot}$), we are sensitive only to
the upper envelope of the trend observed at $z\sim2$.

% conversion b/w IR luminosity (erg/s) and IR solar luminosity (3.84e33)

Three of the \irc \ members are Ultra-Luminous Infrared Galaxies
(ULIRGs; L$_{IR}\geq10^{12}$~L$_{\odot}$; see
Fig.~\ref{fig:sfr_smass}) and the remaining 14 are LIRGs
($10^{11}\leq$L$_{IR}<10^{12}$~L$_{\odot}$).  In stark contrast, of
the $>2000$ galaxies in clusters at $z<1$ studied with wide-field
($R_{proj}\gtrsim1$~Mpc) mid-IR imaging
\citep[e.g.][]{geach:06,saintonge:08,gsmith:10}, only one is a ULIRG
and it lies outside the core of the Bullet Cluster, a well-known
cluster-cluster merger at $z=0.297$ \citep{chung:10}.  The higher
fraction of IR luminous galaxies at $z=1.62$ is likely to be driven by
the overall evolution of the IR luminosity function in clusters
\citep[e.g.][]{bai:09}; however, more deep IR imaging of clusters at
$z>1$ is needed to confirm this trend.

\begin{figure}
\epsscale{1.0}
\plotone{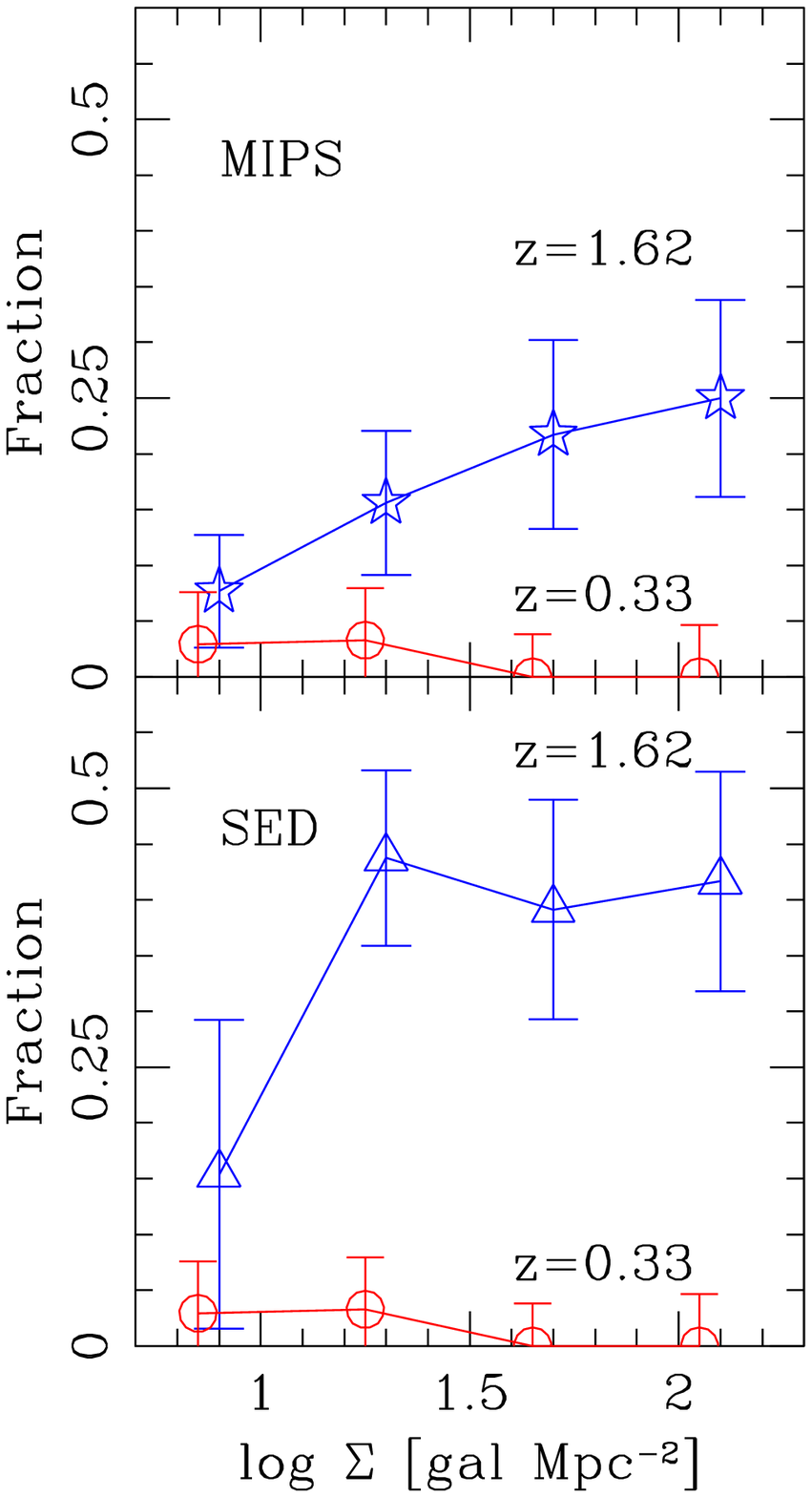}
\caption{Relative fraction of star-forming cluster galaxies vs. local
galaxy density for \mipsmu\ detected members with star formation rates
$\gtrsim40$\msunyr\ ({\it top}; open stars) and members with
SED-derived rates $>$\sfrsed\ ({\it bottom}; open triangles). The open
circles in both panels show the \mipsmu\ detected members in the
massive galaxy cluster CL~1358+62 at $z=0.33$ \citep{tran:09}
and represents the well-established trend documented at $z<1$; 
these circles are offset slightly in $\log\Sigma$ for clarity.
Errorbars are determined from bootstrapping the data in each
bin 1000 times. {\it The fraction of star-forming members is highest
in the regions with the highest galaxy density, a  reversal of the
trend at $z<1$.}  At $z=1.62$, we have reached
the epoch when massive cluster galaxies are still forming a
significant number of new stars.\label{fig:sfr_density}} 
\end{figure}

In our analysis, we assume that the \mipsmu\ sources are dominated by
star formation and do not harbor active galactic nuclei (AGN).  From
the SED fits, we find that two of the \mipsmu\ detected members do
have emission at $8.0\mu$m (IRAC channel 4) that deviates from the
stellar fit, i.e. have a power-law component indicative of an AGN, and
one of these galaxies is the most IR luminous member with
$\log($L$_{IR})$~[L$_{\odot}$]$\sim12.3$ (see
Fig.~\ref{fig:sfr_smass}).  However, studies find that most of the
emission ($\gtrsim70$\%) in ULIRGs is due to star formation
\citep{farrah:08}, thus these members are likely to have both strong
star formation and an AGN component.  Note that at $z=1.62$, the
$7.7\mu$m PAH band lies partly in the \mipsmu\ channel.  Because part
of the IR luminosity is due to star formation, we include both
\mipsmu\ members in our analysis; repeating our analysis without these
two members confirms that our overall results do not change.

In the cluster core ($R_{proj}=0.5$~Mpc), the star formation rate
density from the \mipsmu\ photometry alone is
$\sim1700$\msunyr~Mpc$^{-2}$; we stress that this is likely a lower
limit given the \mipsmu\ imaging cannot detect any members with
SFR$_{IR}\lesssim40$\msunyr, i.e. with star formation rates typical
for IR-detected galaxies in clusters at $z<1$.  Only one other galaxy
cluster at $z=1.46$ has a comparably high star formation rate in its
core \citep{hayashi:10,hilton:10}.  In comparison, studies of
IR-detected galaxies in clusters at $z<1$ find that star-forming
members are strongly segregated at $R_{proj}>0.5$~Mpc
\citep{geach:06,saintonge:08,koyama:08}.

Given the high star formation rate in its core, does \irc \ follow the
well-established trend at $z<1$ of decreasing star-formation with
increasing galaxy density
\citep[e.g.][]{hashimoto:98,ellingson:01,gomez:03}?
Figure~\ref{fig:sfr_density} compares the relative fraction of
star-forming members to passive members as a function of local galaxy
density ($\Sigma$) which is defined by distance to the 10th nearest
neighbor \citep{dressler:80}.  We use star-formation rates derived
from the \mipsmu\ imaging as well as from the SED fitting because the
two independent star formation tracers complement each other and
provide an important check of our results: The \mipsmu\ imaging is a
robust measure of the dust-obscured star formation but only detects
the most active members ($\gtrsim40$\msunyr) while the SED fitting is
sensitive to lower levels of unobscured activity ($\gtrsim$\sfrsed).

Both star formation tracers confirm that {\it the relative fraction of
active members is highest in the regions of highest galaxy density}
(Fig.~\ref{fig:sfr_density}), i.e. exactly opposite to that observed
in clusters at $z<1$.  A Spearman rank test supports with $>97$\%
confidence ($>2\sigma$ significance) that the relative fraction of IR
luminous members increases with increasing galaxy density from
$\sim8$\% at $\Sigma\sim10$~gal~Mpc$^{-2}$ to $\sim25$\% at
$\Sigma\gtrsim100$~gal~Mpc$^{-2}$.  We stress that excluding the two
candidate AGN does not change the trend, and the robustness of this
result is underscored by the fact that we see the same trend using the
SED-derived rates.  While studies of galaxies in the field at $z\sim1$
($\Sigma<10$~gal~Mpc$^{-2}$) find that the star formation-density
relation is beginning to turn over at this epoch
\citep{elbaz:07,cooper:08}, this is the first confirmation of such a
reversal in the significantly higher density environment of galaxy
clusters.

The measured IR luminosities correspond to specific star formation
rates (star formation rate divided by stellar mass; SSFR) per Gyr of
$\sim1-20$: these active members can more than double their stellar
masses in the next Gyr (by $z\sim1.2$).  However, to reproduce the
relatively homogeneous stellar ages measured in massive cluster
galaxies at $z<1$ \citep[e.g.][]{blakeslee:06,tran:07,mei:09}, these
IR luminous members cannot maintain such a high SSFR for even a Gyr.
The current star formation must be quenched rapidly and any later
bursts of activity cannot add a substantial amount of new stars, at
least not in the massive members
($\log($M$_{\ast})$~[M$_{\odot}$]$>10.6$) that must populate a
well-defined red sequence by $z\sim0.8$.

%%%%%%%%%%%%%%%%%%%%%%%%%%%%%%%%%%%%%%%%%%%%%%%%%%%%%%%%%%%%%%%%%%%%
\section{Conclusions}

We measure the rest-frame colors (dust-corrected), IR luminosities,
star formation rates, and stellar masses of galaxies in \irc, a
Spitzer-selected cluster at $z=1.62$, by fitting spectral energy
distributions to photometry in 10 bands
($0.4\mu$m$<\lambda_{obs}<8\mu$m) and with deep \mipsmu\ imaging.  The
cluster sample \citep[Pap10;][]{tanaka:10} is composed of 12
spectroscopically confirmed members and 80 members selected from
photometric redshifts measured using EAZY \citep{brammer:08}; all
members are within $R_{proj}\lesssim1$~Mpc of the massive cluster
galaxy located at the peak of the X-ray emission.

The 92 cluster members have a color-stellar mass distribution that is
surprisingly similar to that observed in field galaxies at $z\sim2$.
When corrected for dust, the cluster members define a strong blue
sequence and span a range in color, indicating a substantial amount of
recent and ongoing star formation in the cluster core.  This dramatic
level of activity is underscored by the 17 members detected at
\mipsmu.  In the cluster core ($R_{proj}<0.5$~Mpc), the star formation
rate density from the IR luminous members alone is
$\sim1700$\msunyr~Mpc$^{-2}$; the true value is likely to be higher
given that we only include members with SFR$_{IR}\gtrsim40$\msunyr.
These IR luminous members also follow the same trend of increasing
star formation with stellar mass that is observed in the field at
$z\sim2$.

We discover the striking result that the the relative fraction of
star-forming galaxies increases with increasing local galaxy density
in \irc, a reversal of the well-established trend at lower redshifts
and in line with recent work at $z\sim1.46$ that suggests enhanced
star formation in cluster cores.  Measurements using star formation
rates derived from the \mipsmu\ imaging and from the SED fitting
provide independent confirmation that the relative fraction of
star-forming galaxies triples from the lowest to highest density
regions.  By pushing into the redshift desert ($z\gtrsim1.6$), we are
able to reach the epoch when massive cluster galaxies are still
forming a significant number of their stars.

\acknowledgements

KT acknowledges generous support from the Swiss National Science
Foundation (grant PP002-110576).  JSD acknowledges the support of the
Royal Society via a Wolfson Research Merit award, and also the support
of the European Research Council via the award of an Advanced
Grant. This work is based in part on data obtained as part of the
UKIRT Infrared Deep Sky Survey.  A portion of the Magellan telescope
time was granted by NOAO, through the Telescope System Instrumentation
Program (TSIP; funded by NSF).

%\bibliographystyle{/Users/vy/aastex/apj}
%\bibliography{/Users/vy/aastex/tran}

%\end{document}

\end{document}